\documentclass[prl, twocolumn]{revtex4}
\usepackage{amsmath}
\usepackage{graphicx}
\usepackage{amsfonts}
\usepackage{amssymb}

\begin{document}

\title{Distinct pseudogap and superconducting state quasiparticle relaxation dynamics in near-optimally doped SmFeAsO$_{0.8}$F$_{0.2}$ single crystals}
\author{T. Mertelj, V.V.Kabanov, C.Gadermaier, J.Karpinski and D. Mihailovic }
\affiliation{$^{1}$Complex Matter Dept., Jozef Stefan Institute, Jamova 39, Ljubljana,
SI-1000, Ljubljana, Slovenia }
\affiliation{$^{2}$Laboratory for Solid State Physics, ETH Z\"{u}rich, 8093 Z\"{u}rich, Switzerland}
\date{\today}

\begin{abstract}
We use femtosecond spectroscopy to investigate the quasiparticle relaxation and low-energy electronic structure in a near-optimally doped pnictide superconductor with $T_{\mathrm{c}}=49.5$ K. Multiple relaxation processes are evident, with distinct superconducting state quasiparticle (QP) recombination dynamics exhibiting a $T$-dependent superconducting (SC) gap, and a clear  "pseudogap" (PG) -like feature with an onset around 200K indicating the existence of a temperature-independent gap of magnitude $\Delta_{\mathrm{PG}}= 61 \pm 9$ meV above $T_{\mathrm{c}}$. Both the SC and PG components show saturation as a function of fluence  with distinct saturation thresholds $\sim0.15 \mu$J/cm$^{2}$ and 1.5 $\mu$J/cm$^{2}$, respectively.
\end{abstract}

\maketitle

Recently, the discovery of high-temperature superconductivity in iron-based pnictides \cite{Kamihara,Ren} has attracted a great deal of attention, partly because of their high critical temperatures, but more fundamentally because they appear to have some similarities and important differences compared to cuprate superconductors, which raises the question of the superconductivity mechanism. Particularly interesting is the proximity of the superconducting phase to a structural transition between tetragonal and orthorhombic symmetry and a magnetically ordered state in both systems, as well as the appearance of phase separation - either intrinsic or chemical\cite{uSR}. A further similarity arises from the fact that Fe$^{2+}$  as well as Cu$^{2+}$ are both Jahn-Teller ions, which may lead to structural instability under conditions of charge disproportionation.  On the other hand, the Fe-As planes are strongly corrugated, suggesting strictly 2D physics is not relevant in this case.
In spite of the rapid emergence of a large number of publications on different aspects of pnictide superconductor physics, many works so far have raised more questions than answers. Particularly pertinent are the questions regarding the existence of a pseudogap precursor state, which is believed by many to be essential for high-temperature superconductivity and is - according to many experiments and theories - attributable to pre-formed pairs above $T_{\mathrm{c}}$.\cite{akm96}

Time resolved spectroscopy has been very instrumental in elucidating the nature of the electronic excitations in superconductors, particularly cuprates, by virtue of the fact that different components in the low-energy excitation spectrum could be distinguished by their different lifetimes \cite{Stevens,  Kabanov,Ours,Gedik,nasiRT,kaindl,MgB2,Bianchi,Schneider}. Moreover, the relaxation kinetics can give us valuable information on the mechanism for superconductivity \cite{Kusar2}. Extensive and systematic experiments on cuprates have also given information on the behaviour of the pseudogap for charge excitations, complementing the information obtained on spin excitations from NMR and other spectroscopies \cite{Ours}. In these experiments, electrons and holes are first excited by laser pulses into states far away from the Fermi energy. The photoexcited charge carriers then relax by carrier-carrier scattering and carrier-phonon collisions to quasiparticle states near the Fermi energy within 10-100 fs. Their subsequent relaxation slows down due to the presence of gaps in the density of states at low energy and can be monitored by measurement of the photoinduced changes of the reflectivity.

In this work we present the first time-resolved femtosecond spectroscopy study of optimally doped SmFeAsO$_{0.8}$F$_{0.2}$ single crystals with the aim of elucidating the electronic structure and to obtain detailed information about the QP dynamics. The present experiments were performed using the standard pump-probe technique, with 50 fs pump pulses at 400 nm and probe pulses at 800 nm. The crystals were flux grown at ETH in Zurich and were approximately 120 x 80 $\mu$m in size. We present the transient reflectance response as a function of temperature and pump intensity. In spite of the very small size of the crystals, we were able to obtain data of sufficiently high quality to make a concise and detailed analysis.

\begin{figure}[htp]
 \begin{center}
   \includegraphics[angle=-90, width=0.4\textwidth]{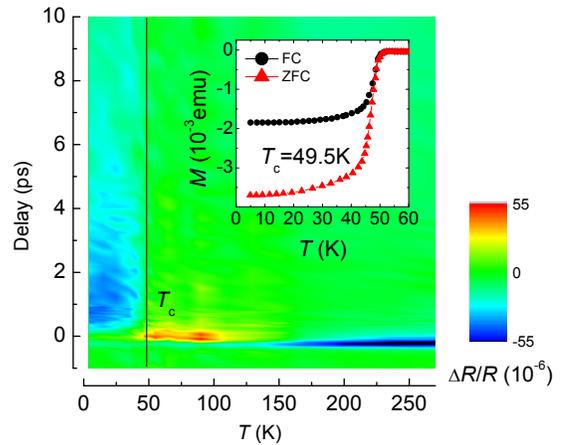} 
   \caption{Temperature dependence of the  the low pump-fluence photoinduced reflectivity, $\Delta R/R$. Below $T_{\mathrm{c}}$ the response of the superconducting state is clearly seen.  In the inset the temperature dependence of the magnetization in the superconducting state indicates high quality of the single crystals.}
      \end{center}
\end{figure}

In Fig. 1 we show the time and temperature dependence of the photoinduced reflectivity  $\Delta R/R$ at low pump-fluences. We can immediately identify three temperature regions: Above $\sim 200$ K, we observe a single dominant negative response (A) which has a relaxation time $\tau_{\mathrm{A}} = 0.3$ ps. As the temperature is lowered, a positive, slightly longer-lived response (B) gradually appears, reaching a temperature independent plateau below 100 K which remains down to $T_{\mathrm{c}}$.  Below $T_{\mathrm{c}}$, an additional third, much longer-lived negative superconducting response (C) appears.  

The raw data below and above $T_{c}$ at low fluence $\mathcal{F}=3 \mu$J/cm$^2$ are shown superimposed on each other in Figures 2a) and b) respectively. 
Since it is not immediately obvious from either figure whether the three signals are superimposed on top of each other, or evolve into each other, we have additionally performed pump laser fluence-dependence measurements at different temperatures in order to disentangle the three components. We find that signal A is linear with $\mathcal{F}$ (see insert to Fig. 2), while B and C are strongly nonlinear with very different non-linearity thresholds. This, together with the strong temperature dependence of signals B and C gives us enough information to uniquely disentangle all three components at low temperatures. 

We begin the analysis by rescaling the  signal obtained at high fluence (shown by the thick dashed traces in Fig. 2) to the fluence at which the low-fluence data was taken and then subtracting it from all the low-fluence scans at all temperatures. (For greater accuracy, we also take into account the additional weak $T$-dependence of the subtracted response at longer delays by measuring the high-fluence response at 150 $\mu$J/cm$^{2}$ at each temperature separately.) Above $T_{\mathrm{c}}$ the result of the subtraction is a $single$ positive exponential decaying response component B shown as a function of $\mathcal{F}$ and temperature in Figures 3b) and 4b) respectively. Exponential fits to the decay of component B at 75 K for different $\mathcal{F}$ are also shown in Fig. 3b).
From Fig. 4b), we see that component B - which we assign to a pseudogap (PG) - starts to appear above 200 K and increases in amplitude with decreasing $T$, reaching a maximum around 100K. It then remains relatively constant in amplitude up to $T_{\mathrm{c}}$. It clearly remains present below $T_{c}$, as seen 
in Fig. 4b), where we have also included the trace at 4.2 K with only signal A subtracted, to unambiguously show that the (negative) C and (positive) B components are simultaneously present. The B  component shows saturation as a function of laser fluence, near $\mathcal{F}=40 \mu$J/cm$^{2}$ (see insert to Fig. 3b).
Its relaxation time is weakly dependent on fluence, with $\tau_{\mathrm{B}}\sim 0.4$ ps at low $\mathcal{F}$, increasing to $\sim 0.5$ ps at saturation.

In Fig.4a) we present the response in the superconducting state, where signals A and B have been subtracted. The superconducting response appears unambiguously below $T_{c}$, has a negative sign, a risetime, $\tau_{\mathrm{rise}}= 0.2$ ps, and a relaxation time of $\tau_{\mathrm{SC}} \sim $ 4 ps at the lowest fluence. It also exhibits a saturation at a relatively low laser power density of $\mathcal{F}_{\mathrm{sat}}  \approx 4 \mu$J/cm$^{2}$ (see insert to Figure 3a)). The risetime increases with increasing fluence reaching $\sim$0.5 ps at the saturation fluence. 
The saturation thresholds, where a departure from the linear $\mathcal{F}$-dependence is observed \footnote{Due to the noise we were not able to reach the region where the SC response is linear in $\mathcal{F}$ so we estimate $\mathcal{F}_{\mathrm{SC}}$ from the ratio of the saturation fluences.},  are $\mathcal{F}_{\mathrm{SC}}\approx 0.15\mu$J/cm$^2$ and $\mathcal{F}_{\mathrm{PG}}= 1.5 \mu$J/cm$^2$  respectively.

\begin{figure}[htp]
 \begin{center}
   \includegraphics[width=0.30\textwidth]{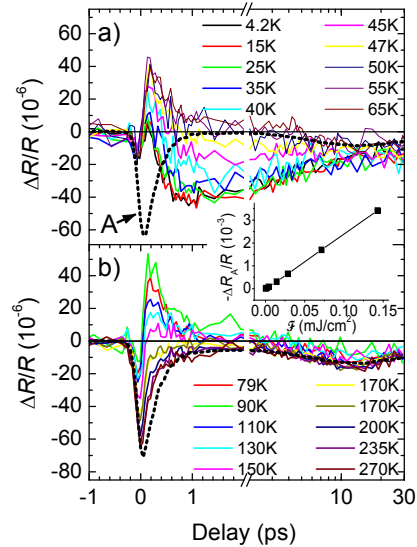} 
   \caption{a) The low-fluence photoinduced reflectivity at different $T$ in the temperature region where superconductivity appears. The thin lines represent scans above $T_{\mathrm{c}}$ while the thick dotted trace shows the high fluence trace at 4.2K linearly scaled to the fluence of the other traces.  b) The photoinduced reflectivity response above $T_{\mathrm{c}}$. The thick dotted trace shows the high fluence response at 200K linearly scaled to the fluence of the other traces. The insert in b) shows the dependence of the T-independent component $\Delta R_{\mathrm{A}}/R$ as a function of $\mathcal{F}$ at 270K.}
      \end{center}
\end{figure}

 \begin{figure}[htp]
 \begin{center}
   \includegraphics[width=0.30\textwidth]{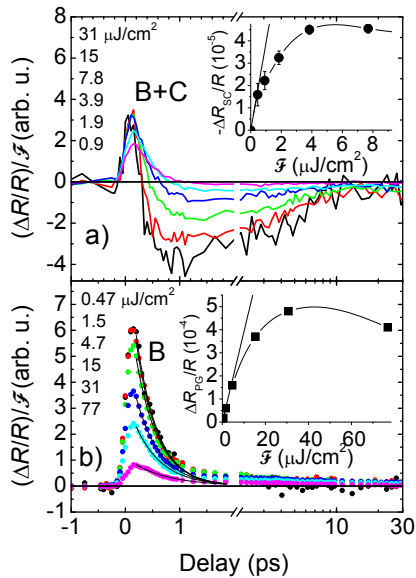} 
   \caption{The photoinduced reflectivity a) at $T=4.2K$ and b) at $T=75$K for different laser fluences $\mathcal{F}$. (The legend is in the same order as the traces). Component A was subtracted in both plots. The inserts show the amplitude of the SC and PG components as a function of pump laser fluence $\mathcal{F}$.  The thin lines are exponential fits.}
   \end{center}
\end{figure}

The amplitudes and relaxation times for the SC and PG responses as a function of temperature are summarized in Fig. 5.
We can understand the $T$-dependence of the $\Delta R_{\mathrm{SC}}/R_{\mathrm{SC}}$ by considering the
difference in reflectivity between the superconducting state and the normal state created by the high excitation conditions. The induced change in reflectivity for fluences which destruct the
SC state is proportional to $\sigma
_{1}^{n}-\sigma _{1}^{s}$, where $\sigma _{1}^{n}$ and $\sigma _{1}^{s}$ are
real parts of the complex conductivity in the normal and superconducting
states, respectively. By using the high frequency limit of the Mattis-Bardeen formula\cite{MattisBardeen58}, $\hbar\omega\gg\Delta_0$, we obtain\cite{Kusar2}:
\begin{equation}
\frac{\Delta R_{\mathrm{SC}}(T)}{R_{\mathrm{SC}}} \varpropto \frac{2\Delta (T)}{\hbar \omega }\ln \left( \frac{%
1.47\hbar \omega }{\Delta (T)}\right),  \label{SatTdep}
\end{equation}%
where $\hbar \omega $ is the photon energy, $\Delta (T)=\Delta _{0}\left( 1-\left( T/T_{c}\right)
^{2}\right) $ and $\Delta _{0}$ is the SC gap at 0 K. Assuming the gap ratio $2\Delta_0/k_BT_{\mathrm{c}} = 3.68$ from tunnelling data on SmFeAs$_{0.85}$F$_{0.15}$  \cite{Chen},  and using $2\Delta _{0} = 16 $ meV for our case we obtain very good agreement between Eq.(%
\ref{SatTdep}) and the data above $\mathcal{F}_{\mathrm{sat}}$ (see fit in Fig. 4b).
 For $%
\mathcal{F}>\mathcal{F}_{\mathrm{sat}}$, the $T$-dependence of $\Delta R_{\mathrm{SC}}/R$
does not depend on $\mathcal{F}$, since full destruction of the SC state is achieved at all
$T<T_{\mathrm{c}}$. Near the threshold, for $\mathcal{F}=3 \mu $J/cm$^{2}$, only partial
destruction is evident below $\sim \frac{2}{3} T_{\mathrm{c}}$ and the amplitude $\Delta R_{\mathrm{SC}}/R$ merges
with the high fluence data as $T\rightarrow T_{c}$. Unfortunately the small sample size and imperfect surface quality prevented us from obtaining high quality data in the low excitation limit below the saturation threshold\cite{nasiRT}.

 \begin{figure}[htp]
 \begin{center}
   \includegraphics[width=0.32\textwidth]{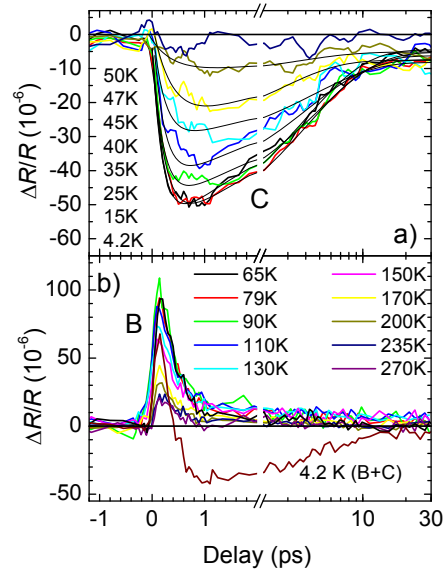} 
   \end{center}
   \caption{a) The photoinduced reflectivity due to the superconducting state response at different $T$ below $T_{\mathrm{c}}$. To obtain the traces the averaged response above $T_{\mathrm{c}}$ (thin traces in Fig. 2a)) was subtracted and the result was smoothed to reduce noise. The thin lines represent a two-exponential fit. b) The photoinduced reflectivity response above $T_{\mathrm{c}}$ showing the PG response. The trace at 4.2 K shows the raw data with only component A subtracted showing unambiguously the simultaneous presence of the SC and the PG components below $T_{\mathrm{c}}$.}
\end{figure}

Turning now to the $T$-dependence of the PG component (Fig. 5b), assuming for simplicity that a $T$-independent gap can be used to describe the PG (as was commonly done in cuprates\cite{Kabanov,Ours,MgB2}), we use Kabanov's formula for the photoinduced reflectance amplitude which describes the QP density in the excited state of a gapped system in a low excitation limit\footnote{At $\mathcal{F}=3 \mu $J/cm$^{2}$ the PG component is just slightly above the linear-response fluence region, $\mathcal{F}< 1.5 \mu$J/cm$^2$, justifying use of the low excitation limit.} under bottleneck conditions\cite{Kabanov,nasiRT}:
\begin{equation}
\frac{\Delta R_{\mathrm{PG}}}{R} \propto [1+B \exp(-\frac{\Delta_{\mathrm{PG}}}{k_{\mathrm{B}} T})]^{-1}.
\end{equation}
Here $B=2\nu /(N(E) \hbar \Omega_{\mathrm{c}})$, $N(E)$ is the density of electronic states at $E_{\mathrm{F}}$, $\nu$ is the number of bosons involved in the relaxation process across the PG, and $\Omega_{\mathrm{c}}$ is the cut-off frequency of the bosonic spectrum.
Using $N(E)\simeq 2 $eV$^{-1}$cell$^{-1}$spin$^{-1}$, and $\Omega_{\mathrm{c}} \simeq 60 $ meV from ref. \cite{Singh}, a fit to the data in Fig. 5b) gives an effective pseudogap magnitude of $\Delta_{\mathrm{PG}}= 61 \pm 9$ meV and $\nu=4.4 \pm 1.5$. We see that in spite of the fact that the model is relatively simple, it describes the PG temperature dependence relatively well.

\begin{figure}[htp]
\begin{center}
   \includegraphics[angle=-90, width=0.45\textwidth]{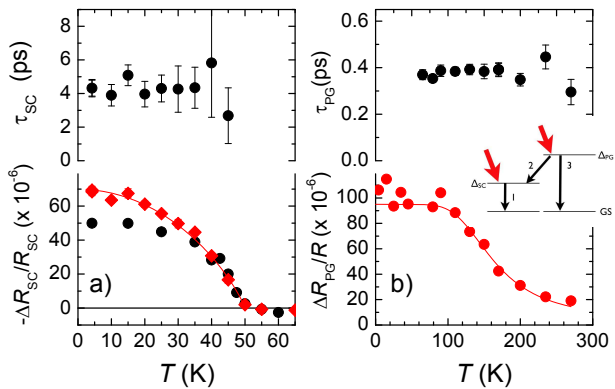} 
   \end{center}
   \caption{a) The superconducting state relaxation time, $\tau_{\mathrm{SC}}$, and amplitude of the photoinduced reflectivity $\Delta R_{\mathrm{SC}}/R_{\mathrm{SC}}$ in the superconducting state. $\Delta R_{\mathrm{SC}}/R_{\mathrm{SC}}$ is shown as a function of $T$ for two laser fluences: 3 $\mu J/cm^2$ (circles) and 15 $\mu J/cm^2$ (diamonds). The line is a fit using the  expression (Eq. 1) for the photoinduced reflectivity in the strong excitation limit.  b) The amplitude and relaxation time, $\tau_{\mathrm{PG}}$, of the "pseudogap" component above $T_{\mathrm{c}}$ for $\mathcal{F}=3$ $\mu J/cm^2$. The fit to the data uses model\cite{Kabanov} with a $T$-independent gap. The insert shows the possible recombination pathways across $\Delta_{\mathrm{SC}}$ and $\Delta_{\mathrm{PG}}$. The red arrows indicate the initial very fast relaxation. The PG relaxation may proceed via 2 and then 1 or directly to the ground state 3.}
\end{figure}

Two important conclusions can be reached immediately from the data. The first is that the SC state QP recombination dynamics in SmFeAs$_{0.8}$F$_{0.2}$ follows the general behavior observed in systems exhibiting a $T$-dependent gap\cite{Kusar2}. The second conclusion is that the system exhibits a pseudogap in the low-energy spectrum, which appears in our signal below  $T^* \sim 200$ K and is clearly present in the SC state. Its magnitude is $\Delta_{\mathrm{PG}} = 61 \pm 9$ meV, where the error signifies the dependence on the model parameter $\nu$.  The value of $\nu \sim 4.4 \pm 1.5$ used here to some extent indicates the strength of coupling to the phonons, assuming of course that the bosons involved in the relaxation process are phonons. 

The question immediately arises on the origin of the PG. One possibility is that it corresponds to a higher lying excited state, and that the relaxation is a process governed by a relaxation cascade shown in the insert to Fig. 5 as $2 \rightarrow 1$, which proceeds from the PG state via the QP state to the ground state (GS).  (In cuprates, a similar question has arisen \cite{Bianchi}). One indication against the cascade process is that at the lowest fluences the risetime of the SC component $\tau_{\mathrm{rise}}=0.2$ ps appears shorter than $\tau_{\mathrm{PG}}=0.4$ ps just above $T_{\mathrm{c}}$. However, below $T_{\mathrm{c}}$, the experimental error, due to the overlap of the two signals, prevented us to clearly identify two distinct relaxation times, so the cascade can not be entirely ruled out.

If the cascade is not responsible for the observed relaxation kinetics, the question of the origin of the two coexisting relaxation processes arises. A possible explanation is the coexistence of the two states as a result of  either microscopic or mesoscopic phase separation in real space, such as was discussed in cuprates\cite{Ours}. Indeed structural phase coexistence of tetragonal and orthorhombic phases can be deduced from the appearance of split or asymmetric Bragg diffraction peaks in high resolution X-ray diffraction experiments on SmFeAsO$_{1-x}$F$_{x}$, even for $x$ above 0.15 \cite{Margadonna}. 

Alternatively $\Delta_{\mathrm{SC}}$ and $\Delta_{\mathrm{PG}}$ may appear in different regions of the Fermi surface, as suggested by band structure calculations which predict a flat band and two nearly degenerate hole-like bands crossing the FS near ${\bf \Gamma}$ point, and electron-like bands near the ${\bf M}$ point in the Brillouin zone respectively \cite{Singh}. Angle-resolved photoemission spectroscopy may eventually be able to resolve the two gaps, provided that surface-related experimental problems do not prove to be insurmountable. 

We conclude that on the basis of the present experiments that the low-energy electronic structure of the pnictide cannot be described in terms of a simple homogeneous metal with a single SC gap. The presence of the pseudogap in our single crystal data is quite unambiguous and is much more strongly expressed than in tunneling experiments on polycrystalline samples by Pan et al.\cite{Pan}, or $^{19}F$ NMR relaxation time measurements \cite{NMR}, (also measured in polycrystalline samples) where only a relatively weak monotonic drop of $1/T_1T$ is observed.  The observations strongly suggest a similarity with the low-energy pseudogap structure of the cuprate superconductors and is described phenomenologically quite well by a temperature-independent gap in the low energy spectrum of 61 meV. Moreover, the QP recombination rates associated with the SC gap and the PG are very similar to La$_{2-x}$Sr$_{x}$CuO$_{4}$ and YBa$_{2}$Cu$_{3}$O$_{7-\delta}$ \cite{Ours} for example, but nearly two orders of magnitude shorter than in MgB$_{2}$, implying that the  QP recombination mechanism in the SmFeAsO$_{0.8}$F$_{0.2}$ is similar to cuprates, and very different to MgB$_{2}$.

We wish to acknowledge valuable discussions with A.S.Alexandrov.

\end{document}